\documentclass[usenatbib]{mn2e}
\usepackage{amsmath}
\usepackage{natbib}
\usepackage{enumitem}
\usepackage{graphicx}
\usepackage{times}
\begin{document}

\date{Accepted 2013 September 24.  Received 2013 September 24; in original form 2013 June 5}

\title{A Dynamical Analysis of the Corona Borealis Supercluster}

\author[Merida Batiste \& David J. Batuski]{Merida Batiste\thanks{E-mail:merida.batiste@umit.maine.edu} \& David J. Batuski\thanks{E-mail:david.batuski@umit.maine.edu}\\Department of Physics and Astronomy, University of Maine, 120 Bennett Hall, Orono, ME 04469, USA}

\maketitle

\begin{abstract}
Using data from the Sloan Digital Sky Survey we assess the current dynamical state of the Corona Borealis Supercluster (CSC), a highly dense and compact supercluster at $z\approx0.07$. The Fundamental Plane relation is used to determine redshift independent distances to six clusters in the densest region of the supercluster, with mean accuracy in the relative distance estimates of $4$ per cent. Peculiar velocities determined from these distance estimates indicate that the clusters have broken from the Hubble Flow, suggesting that the CSC likely contains two regions that have reached turnaround and are currently undergoing gravitational collapse. These results provide the strongest observational evidence to date that the CSC is a bound system similar to the much more extensive Shapley Supercluster, which is the most extensive confirmed bound supercluster yet identified in the Universe. When compared with simulations of the CSC our results require substantially more mass than is contained within the clusters, possibly indicating a significant inter-cluster dark matter component. In order to facilitate comparison with studies for which spectroscopic data are not available, an alternative analysis of the dynamics is made using the Kormendy relation as a distance indicator. The results are generally consistent with those of the Fundamental Plane and suggest similar global dynamics, but we find that the relatively sparse sampling of the clusters makes the Kormendy relation less reliable overall and more susceptible to small systematic differences between the cluster samples.

\end{abstract}


\section{Introduction}
Superclusters of galaxies are the most massive and extended galactic systems in the Universe. They can contain as many as 25 clusters and groups of galaxies spanning regions $30 - 100$ Mpc across, with masses larger than $10^{15} h^{-1} M_{\odot}$. Observations on the largest scales have shown that many such systems exist and that they are rarely isolated from one another, instead being connected by filaments and sheets of galaxies, which surround large voids \citep{Einasto80}. Superclusters can, then, most generally be described as the high density regions that form in the interstices between voids, and some limiting overdensity is usually chosen to define the boundaries of these systems.
Dynamical, photometric and morphological studies of superclusters are of tremendous interest because they can be used to test theories of formation and evolution of structure in a range of environments \citep{Einasto80,Bardelli93,Baldi01}. In the densest superclusters these studies may also be used to place constraints on the large scale distribution of dark matter \citep{Quintana95}, to test for the effects of dark energy on the formation and evolution of structure, and to place limits on the extent of bound structure in the Universe \citep{Araya09}. 

Early studies \citep{Rood76} of superclusters showed that the internal dynamics are, in general, dominated by Hubble flow, but the Shapley Supercluster (hereafter SSC) is a rare counter-example to this. First noted by \citet{Shapley30}, the SSC, located at RA = $13^{h} 25^{m}$ and Dec = $-30^{\circ}$, at a redshift of $z \approx 0.05$, consists of $\sim25$ clusters and groups spanning a region of $\sim 100$ Mpc, making it the most massive supercluster complex identified in the Universe. It has been intensively studied in a range of wavelengths (e.g. \cite{Bardelli93,Ettori97,Reisenegger02,Proust06}, and references therein) and has been shown to contain significant bound structure, with a central core that is in the final stages of collapse \citep{Reisenegger00}. To date the SSC is the only confirmed bound supercluster in the Universe (for the purposes of the current paper we define a supercluster as containing no fewer than $3$ clusters, or groups, of galaxies). However, during the last two decades several similarly dense superclusters have been identified as possible bound structures, prompting the suggestion of a more physically motivated definition of superclusters as the largest gravitationally bound structures in the Universe \citep{Araya09}.

The Corona Borealis Supercluster is another of the densest, and most compact, superclusters in the local universe ($z\approx0.07$) and several authors have suggested that it should have sufficient cluster density to be gravitationally bound \citep{Postman88,small3,Kopylova98}. It was included by \citet{Abell58} in his catalog of second order clusters and first noted by \citet{Shane54}, who initially identified 12 member clusters in a $6^{\circ} \times 6^{\circ}$ region before later using brightest cluster galaxies to show that there were actually two components viewed in projection. The background component consists of five Abell clusters, three of which are at $z\approx0.11$ while two are even more distant \citep{small1}. The foreground component, at $z\approx 0.07$, contains Abell clusters 2092, 2089, 2079, 2067, 2065 and 2061 and is what we now refer to as the Corona Borealis Supercluster (hereafter CSC). \citet{Postman88} performed the first dynamical analysis of this region, using $182$ galaxy redshifts to make virial mass estimates of each of these six clusters. They found a mass for the supercluster of $8.2 \times 10^{15} h^{-1} M_{\odot}$, concluding that this was likely sufficient to bind the system with peculiar velocities $\leq2200\;km\; s^{-1}$. The Norris Survey \citep{small1,small2,small3} expanded on this work, increasing the number of galaxy redshifts for the six clusters to $528$ and performing N-body simulations to test mass estimators and assess the likelihood of the CSC being bound. By treating the six clusters in the core as a single virialized system, they estimated a mass of at least $3 \times 10^{16}h^{-1}\: M_{\odot}$, concluding that the system was bound and had likely reached turnaround. \citet{Kopylova98} took a slightly different approach, similar to that presented in the current paper, using photometry to determine the peculiar velocities of eight clusters (including 2019 and 2124 in addition to those of \citet{Postman88}) in the supercluster. The Kormendy relation was used to determine redshift independent distances and derive peculiar velocities for each of the clusters. Their results showed significant peculiar velocities for all eight clusters, leading them to define a rapidly collapsing core containing five clusters; 2089, 2092, 2065, 2067 and 2061. 

More recently the work of \citet{small3} has been revisited by \citet{Pearson13}, who correctly contend that a supercluster is far from being a virialised system, so any application of the Virial Theorem is unlikely to be valid. Instead, \citet{Pearson13} have performed N-body simulations of the CSC using the most accurate mass estimates available for the clusters (and assuming negligible inter-cluster mass), concluding that there is very little likelihood that any part of the structure is bound. The results of \cite{Kopylova98} likewise warrant more extensive analysis, since their derived peculiar velocities would require a higher mass to bind the system than that estimated by either Postman et al. or Small et al., and their cluster distance estimates involved fewer than 10 galaxies in all but one cluster, suggesting that the derived peculiar velocities must ultimately be dominated by error. In this paper we aim to improve on these results. 

The paper is structured as follows: in \S\ref{sec:SDI} we describe the Fundamental Plane and how it can be used to obtain accurate cluster distance estimates. In \S\ref{sec:data} we describe the data, explaining our selection criteria and the corrections to photometry. In \S\ref{sec:fitting} we describe our fitting method and present the best fitting Fundamental Plane for the individual clusters, as well as for the supercluster. In \S\ref{sec:dist} we present our cluster distance estimates, with associated errors, and the derived cluster peculiar velocities. In \S\ref{sec:systematics} we discuss possible sources of bias and systematic differences between the cluster samples which might contribute to uncertainty in the distance estimates. In \S\ref{sec:KR} we present an alternative analysis using the Kormendy relation as a secondary distance indicator. In \S\ref{sec:disc} we discuss the physical significance of the results and how they can be used to describe the current dynamical state of the CSC, as well as their implications in light of recent simulations of the CSC. Finally we draw conclusions in \S\ref{sec:conc}. Throughout this paper we adopt a standard $\Lambda$CDM cosmology with $H_{0}=100\:h\:km\:s^{-1}\:Mpc^{-1}$ and $q_{0}=-0.55$. 
\section{The Fundamental Plane}
\label{sec:SDI}
It is well established that early-type galaxies (ETGs) can be described by three structural and kinematic parameters: line of sight velocity dispersion, $\sigma$, effective radius, $r_{e}$, which contains half the total light of a galaxy, and mean effective surface brightness, $\langle\mu\rangle_{e}$, which is given by:
\begin{equation}
\langle\mu\rangle_{e} \propto -2.5\:log\left(\frac{L}{2\pi r_{e}^{2}}\right)
\end{equation}
where $L$ is the luminosity. Assuming that all ETGs are virialized systems we would expect correlations among these observables as a direct consequence of the Virial Theorem, which gives:
\begin{equation}
v^{2}\propto\frac{GM}{R} \propto 2 \pi\: G\left(\frac{M}{L}\right)R\left(\dfrac{L}{2 \pi R^{2}}\right)
\end{equation}
If we assume a constant mass-to-light ratio, as well as structural symmetry and isotropic velocities such that $v^{2}\propto\sigma^{2}$ and $R\propto r_{e}$, this relation can be rewritten as:
\begin{equation} \label{eq:FP}
log\:r_{e}= a\:log\:\sigma + b\:\langle\mu\rangle_{e} + c
\end{equation} 
for which the Virial Theorem predicts $a=2$ and $b=0.4$. Equation \ref{eq:FP} is known as the Fundamental Plane relation (FP) for ETGs \citep{Djorgovski87}, and it incorporates the Faber-Jackson \citep{FJ76}, Kormendy \citep{Kormendy77}, and $D_{n}-\sigma$ \citep{Dressler87} relations, all of which can be seen as projections of this plane. If effective radius is measured in angular units, then the zero point offset $c$ is directly dependent on the distance of a galaxy, and this relation can be used as a redshift independent distance indicator. It is important to note that the FP does not provide a direct measure of distance. Rather it is necessary to calibrate the relation using a galaxy (or, more usually, a cluster of galaxies) with a known distance, so that all distances measured using the FP are inherently relative distances which depend strongly on the calibration used. 

In actuality the observed Fundamental Plane is tilted with respect to that predicted by the Virial Theorem, with ETGs better described by $a\,\approx\,1.2$ and $b\,\approx\,0.3$ \citep{Jorgensen96}. Further, studies of the FP have shown a wide variation in best fitting coefficients, which cannot be explained solely by differences in methods of data reduction, analysis or fitting techniques. This tilt, and the apparent variation in best fitting coefficients, have been extensively studied (e.g. \cite{Gibbons01,Donofrio08,Gargiulo09,Labarb10}) but have so far resisted rigorous explanation. Nonetheless the FP has proven effective as a secondary distance indicator, with intrinsic dispersion similar to that of the Tully-Fisher relation, equivalent to $\sim20$ per cent error in distance to a single galaxy. Since errors in cluster distance estimates scale as $N^{1/2}$, accuracies in cluster distances of better than $5$ per cent (assuming accurate calibration) can be achieved with sample sizes of only 15 galaxies per cluster \citep{Gibbons01}. The advent of large scale surveys, such as the Sloan Digital Sky Survey, make cluster samples of this size a more realistic goal than at any time in the past.

\section{The Data}
\label{sec:data}
\begin{figure}
\begin{center}
\includegraphics{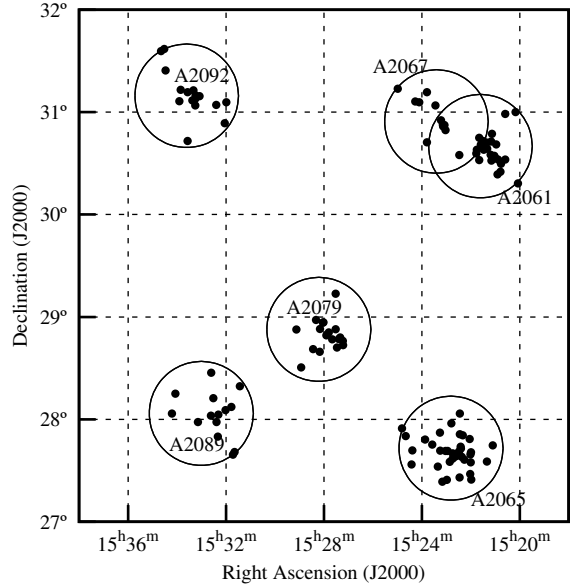}
\caption{Positions on the sky of the six CSC clusters, with all galaxies used for the FP analysis shown. Coordinates for the cluster centres are given in Table \ref{tab:posdata}. The circles each have a radius of 30 arcmin, slightly larger than the Abell radius at the distance of the supercluster.}
\label{fig:sampling}
\end{center}
\end{figure}

We are using a magnitude limited sample of ETGs drawn from the Sloan Digital Sky Survey, Data Release 7 \citep{SDSS7}. Details regarding the photometric and spectroscopic data collection and reduction can be found in \cite{Bernardi03a} and \cite{SDSS2}, and references therein. The initial galaxy samples were obtained by first defining each cluster as having a radius of 30 arcmin, slightly larger than the Abell radius ($\sim25$ arcmin at the distance of the CSC), and redshifts taken from the NASA/IPAC Extragalactic Database (NED). The SDSS catalogue was then queried for each cluster over a range in redshift of $\pm0.015$, within the 30 arcmin radius. For each cluster the galaxy redshifts were then averaged (using the biweight location estimator \citep{Beers}, \S\ref{sec:dist}) and the samples were iteratively clipped at the $3\sigma_{BW}$ level (where $\sigma_{BW}$ is the biweight scale estimator), to reject outliers and non-cluster members. Our final cluster redshifts, shown in Table \ref{tab:posdata}, have been calculated using a biweight average of the galaxy redshifts in each sample once the $3\sigma_{BW}$ clipping (but before the ETG selection criteria, detailed below) has been applied. It should be noted that A2067 and A2061 have been dealt with slightly differently. The proximity of these clusters, both in redshift and on the sky, results in a number of duplicate data points between the two samples (the overlap in the sample is shown in Figure \ref{fig:sampling}). To account for this we apply two additional cuts: the first is based on the distribution of these data on the sky, which shows a clear separation between the clusters. The second cut is based on the redshift distributions for the two cluster samples, both of which are clearly (and similarly) bimodal. Such bimodality in the redshift distribution is not evident in any of the other cluster samples. 

In order to identify ETGs in the SDSS catalogue we have used the selection criteria of \citet{Bernardi03a}, who undertook an extensive investigation of ETGs and their scaling relations using a sample of nearly $9000$ ETGs from the SDSS catalogue. We have made some adjustments and corrections based on more recent investigations of the SDSS data reduction pipeline \citep{Hyde09}, specifically accounting for issues regarding sky subtraction. Briefly, in the language of the SDSS, the selection criteria are as follows:
\begin{enumerate}[leftmargin=0.5cm,label={(\arabic*)},itemsep=3pt]
\item
$FRACDEV\geq\:0.8$
\item
$\frac{b}{a}\geq\:0.6$
\item
$eclass<\:0$
\item
$\sigma\geq\:100\:km\:s^{-1}$
\item
$14.50\:\leq\:r^{*}\leq\:17.45$ 
\end{enumerate}
The SDSS does not give estimates of bulge to disc ratios, so the first three criteria are intended to select for ETGs by rejecting galaxies with a disc component. The first criterion requires that the light profile be best fitted by a de Vaucouleurs $r^{1/4}$ model, characteristic of ETGs and the bulges of spiral galaxies. The second criterion applies an axis ratio cut, since galaxies with $b/a <0.6$ are often partially supported by rotation \citep{BinneyMerri}, and investigation of the FP has shown that contamination with disc galaxies can increase the observed scatter in the relation \citep{Saglia93}. The third criterion uses a PCA classification to remove objects with emission line spectra, characteristic of late-type spiral galaxies. The magnitude limits are based on the completeness of the sample, and match those of \cite{Bernardi03a}. Due to the poor sky subtraction from which the reduced SDSS data are known to suffer we have chosen not use a concentration index, which takes the ratio of petrosian radii containing 90 per cent and 50 per cent of the light of a galaxy respectively, as the measured petrosian radii are particularly susceptible to this problem. 

The spectral resolution of the SDSS means that velocity dispersion estimates lower than $90\: km\: s^{-1}$ are considered unreliable, so we have made a slightly more conservative cut at $100\:km\:s^{-1}$.  Table \ref{tab:posdata} gives the positions, spectroscopic redshifts, and sample sizes for each cluster, and Figure \ref{fig:sampling} shows the positions of the clusters on the sky.

Apparent magnitudes and effective radii derived from the de Vaucouleurs profile fit, which we use in our analysis, have been corrected for seeing, Galactic extinction, and atmospheric extinction. To account for the sky subtraction problems we have made corrections to magnitudes and effective radii based on the analysis of \citet{Hyde09}. We first normalize the de Vaucouleurs radius by the axis ratio:
\begin{equation}
r_{SDSS}\:=\:r_{deV}\:\sqrt{\frac{b}{a}}
\end{equation}
Corrections are then applied to the radius, $r_{SDSS}$, and magnitude, $m_{deV}$, depending on the size of $r_{SDSS}$. Specifically, no correction is applied to the radius if $r_{SDSS}<\:2.0$ arcsec, and none is applied to the magnitude if $r_{SDSS}<\:1.5$ arcsec. Otherwise:
\begin{center}
\begin{eqnarray}
& r_{e}\:=\:r_{SDSS}\:+\:\Delta r_{fit}\\
& m\:=\:m_{deV}\:+\:\Delta m_{fit}
\end{eqnarray}
\end{center}
where:
\begin{eqnarray}
& \Delta r_{fit}\:=\:0.181571\:-\dfrac{r_{SDSS}}{4.5213}\:+\:\left(\dfrac{r_{SDSS}}{3.9165}\right)^{2}\\
& \Delta m_{fit}\:=\:0.024279\:-\:\dfrac{r_{SDSS}}{71.1734}\:-\:\left(\dfrac{r_{SDSS}}{26.5}\right)^{2}
\end{eqnarray}
The mean effective surface brightness, corrected for cosmological dimming and K-corrected following \cite{kcor}, is then given by:
\begin{equation}
\langle\mu\rangle_{e}=m+2.5\:log\left(2 \pi r_{e}^{2}\right)-K(z)-10\:log(1+z) 
\end{equation}
\begin{table*}
\begin{minipage}{150mm}

\caption{Positions of the six clusters on the sky, with right ascension and declination in J2000 coordinates. Column (4) shows spectroscopic redshifts, calculated from a biweight average (\S\ref{sec:dist}) of the galaxy redshifts. Uncertainties in the spectroscopic redshifts are calculated via a boot-strapping procedure. Column (5)  shows the number of galaxies used in the cluster redshift determinations, and column (6) gives the number of ETGs in the FP samples. Column (7) gives mass estimates for each of the clusters. Values for the supercluster are shown in the last line.}
\label{tab:posdata}
\begin{center}
\begin{tabular}{lcccccc}
\hline 
Cluster & RA & Dec & $z_{spec}$ & $N_{spec}$ & $N_{FP}$ & Mass $(10^{15}\:h^{-1}\:M_{\odot})$ \footnote{Mass estimates provided by D.~W.~Pearson (\cite{Pearson13}, and private communication)} \\ 
\hline 
A2092 & $15^{h}33^{m}21^{s}$ & $+31\degr 09^{\prime}32^{\prime\prime}$ & $0.0664\pm 0.0002$ & 59 & 16 & 0.181 \\ 
A2089 & $15^{h}32^{m}45^{s}$ & $+28\degr 03^{\prime}47^{\prime\prime} $ & $0.0737\pm 0.0003$ & 64 & 15 & 0.409 \\ 
A2079 & $15^{h}28^{m}05^{s}$ & $+28\degr52^{\prime}40^{\prime\prime}$ & $0.0659\pm 0.0003$ & 83 & 18 & 0.826 \\ 
A2067 & $15^{h}23^{m}14^{s}$ & $+30\degr54^{\prime}24^{\prime\prime}$ & $0.0740\pm 0.0002$ & 50 & 11 & 0.976 \\ 
A2065 & $15^{h}22^{m}43^{s}$ & $+27\degr43^{\prime}21^{\prime\prime}$ & $0.0717\pm 0.0004$ & 152 & 38 & 1.132 \\ 
A2061 &$15^{h}21^{m}21^{s}$	&$+30\degr40^{\prime}15^{\prime\prime} $ & $0.0788\pm 0.0002$ & 83 & 25 & 1.384 \\
CSC & $15^{h}28^{m}$ & $+30\degr$ & $0.0717\pm 0.003$ &  & 122 & -- \\
\hline
\end{tabular}
\end{center}
\end{minipage}
\end{table*} 
Velocity dispersions are aperture corrected to one eighth the effective radius following \cite{Jorgensen95}:
\begin{equation}
\dfrac{\sigma_{cor}}{\sigma_{SDSS}}=\left(\dfrac{r_{fiber}}{r_{e}/8}\right)^{0.04}
\end{equation}
where $r_{fiber}=1.5$ arcsec, and $r_{e}$ is the corrected effective radius. In what follows, $r_{e}$, $m$, and $\sigma$ have all been corrected following the methods described above.
\section{Fitting the FP}
\label{sec:fitting}
If all cluster early-type galaxies are drawn from the same underlying distribution then, in principle, a single set of coefficients, $a$ and $b$, will provide the best fit for all clusters. In practice, however, there is significant variation in these coefficients in the literature, and currently there is little understanding of the underlying causes. This requires that we find, independently, the coefficients which best describe the six clusters under consideration.
 
It is well established that the best fitting coefficients are strongly affected by the fitting method used, and in large part the choice of method is dictated by the particular type of investigation one wishes to pursue. If these relations are used to investigate the nature of early-type galaxies, or to constrain the underlying physics, then either an inverse fit, which takes $log\:\sigma$ as the dependent variable, or an orthogonal fit, which minimizes residuals orthogonal to the plane, will provide the most unbiased assessments. However, if this relation is to be used as a secondary distance indicator then it is most appropriate to perform a “direct” fit, minimizing residuals in the distance dependent parameter $log\: r_{e}$ (see discussion in \cite{Bernardi03c}).

Whilst the fitting method will certainly affect the coefficients, further bias may be introduced by measurement error, correlations between the measurement errors, and the intrinsic dispersion in the relation. Along with the intrinsic dispersion, correlated errors are of particular concern in this sample since effective radius and apparent magnitude are measured from the same de Vaucouleurs fit to the galaxy light profile. The \textit{Bivariate Correlated Errors and Intrinsic Scatter} (BCES) fit \citep{Akritas} accounts for these biases by employing an average covariance matrix, which accounts for the variance in the errors in all parameters and their mutual correlations, to correct the results of a least-squares fit. If we set $log\:r_{e}\equiv R$ and $log\:\sigma \equiv V$ then the covariance matrix, \textbfss{E}, can be represented as:
\begin{equation*}
\mathbfss{E}=
\begin{pmatrix}
\epsilon_{\mu\mu}^{2} & \epsilon_{R\mu}^{2} & \epsilon_{V\mu}^{2} \\
\epsilon_{R\mu}^{2} & \epsilon_{RR}^{2} & \epsilon_{RV}^{2} \\
\epsilon_{V\mu}^{2} & \epsilon_{RV}^{2} & \epsilon_{VV}^{2}
\end{pmatrix}
\end{equation*}
Following the convention of \citet{Bernardi03a} we denote the error in $log\:r_{deV}$ as $e_{r}$, the error in $m_{deV}$ as $e_{m}$, the error in $log\:\sigma$, before the aperture correction is applied, as $e_{v}$, and the error in the axis ratio as $e_{ab}$. The matrix elements are then given by:
\begin{eqnarray}
& \epsilon_{RR}^{2} = \langle e_{r}^{2} \rangle+\dfrac{\langle e_{ab}^{2}\rangle}{4} \nonumber\\
& \epsilon_{\mu\mu}^{2} = \langle e_{m}^{2}\rangle+25 \epsilon_{RR}^{2} \nonumber\\
& \epsilon_{VV}^{2} = \langle e_{v}^{2} \rangle+(0.04 \epsilon_{RR})^{2} \nonumber\\
& \epsilon_{VR}^{2}= -0.04 \epsilon_{RR}^{2} \nonumber\\
& \epsilon_{V \mu}^{2}= -0.2 \epsilon_{RR}^{2} \nonumber\\
& \epsilon_{\mu R}^{2}=\langle \epsilon_{\mu\mu} \epsilon_{RR} \rangle
\end{eqnarray}
These are similar to those presented by Bernardi et al., with differences arising from that fact that, in this case, we use mean effective surface brightness rather than absolute magnitude. Errors in the spectroscopic redshifts, as well as those in the photometric corrections described in the previous section, have not been included as they are generally an order of magnitude smaller than those included above. 

We fit the Fundamental Plane for each cluster by finding the coefficients that minimize
\begin{equation} \label{eq:FP_residual}
\sum\limits_{i} \left(log\: r_{e_{i}}-a\:log\:\sigma_{i}-b\langle\mu\rangle_{e_{i}}-\:c_{i}\right)^{2}
\end{equation}
 summed over all $N$ galaxies in the cluster. If we set $\rho_{xy}^{2}=\overline{xy}-\overline{x}\;\overline{y}$, where $x,y$ can be $\mu$, $R$ or $V$, then the direct fit coefficients are given by:
\begin{eqnarray}
& a=\dfrac{\rho_{\mu\mu}^{2} \rho_{RV}^{2}-\rho_{\mu R}^{2} \rho_{\mu V}^{2}}{\rho_{\mu\mu}^{2} \rho_{VV}^{2}-\rho_{\mu V}^{4}} \nonumber\\
& b=\dfrac{\rho_{\mu R}^{2} \rho_{VV}^{2}-\rho_{\mu V}^{2} \rho_{RV}^{2}}{\rho_{\mu \mu}^{2} \rho_{VV}^{2}-\rho_{\mu V}^{4}} \nonumber\\
& c_{i}=log\:r_{e_{i}}-a\:log\sigma_{i}-b\:\langle\mu\rangle_{e_{i}}
\end{eqnarray}
The \textit{BCES} corrected coefficients are found by subtracting the appropriate $\epsilon_{xy}^{2}$ from each $\rho_{xy}^{2}$ in the above expressions, providing a largely unbiased estimate of the best fitting coefficients. Offsets $c_{i}$ are calculated for each galaxy, and those values are then averaged to give cluster offsets (\S\ref{sec:dist}). It is worth noting that though this method will not account for any bias due to selection effects, this should not be a concern in using the FP as a distance indicator, since the whole sample will be affected by the selection criteria in the same way.
\begin{table*}
\begin{minipage}{150mm}
\caption{Best fitting FP coefficients for each of the clusters, with associated errors (calculated via a boot-strapping procedure), are shown in columns (2), (3) and (4), with the last row giving the best fits and associated errors for the supercluster. Column (5) gives the rms dispersion in the fits for each cluster, with the last row giving the dispersion in the global fit. Columns (6) and (7) give the cluster distances determined from the FP ($z_{FP}$) and associated error (determined from the rms dispersion in the global fit), respectively, and column (8) gives the resulting peculiar velocities.}
\label{tab:coeffs}
\begin{center}
\begin{tabular}{cccccccc}
\hline 
Cluster & $a$ & $b$ & $c$ & $rms$ & $z_{FP}$ & error ($\%$)& $V_{pec}\: (km\:s^{-1})$ \\
\hline 
A2092 & $1.10\pm 0.26$ & $0.332\pm 0.034$ & $-8.866\pm 0.019$ & $0.063$ & $0.0692$ & 4 & $-796$ \\ 
A2089 & $1.14\pm 0.37$ & $0.360\pm 0.054$ & $-8.845\pm 0.023$ & $0.067$ & $0.0714$ & 4 & $639$ \\ 
A2079 & $0.99\pm 0.21$ & $0.321\pm 0.061$ & $-8.836\pm 0.019$ & $0.081$ & $0.0629$ & 4 & $846$ \\ 
A2067 & $1.18\pm 0.78$ & $0.270\pm 0.190$ & $-8.848\pm 0.033$ & $0.075$ & $0.0724$ & 5 & $448$ \\ 
A2065 & $1.10\pm 0.12$ & $0.359\pm 0.023$ & $-8.859\pm 0.010$ & $0.064$ & $0.0725$ & 3 & $-229$ \\ 
A2061 & $1.50\pm 0.33$ & $0.351\pm 0.034$ & $-8.828\pm 0.016$ & $0.068$ & $0.0719$ & 3 & $1933$ \\
CSC & $1.11\pm 0.07$ & $0.339\pm 0.015$ & $-8.853\pm 0.010$ & $0.069$  & $0.0717$ & -- &--\\
\hline 
\end{tabular}
\end{center}
\end{minipage}
\end{table*} 
The best fitting coefficients for each cluster are shown in Table \ref{tab:coeffs}. There is significant variation from cluster to cluster, but in order to use the FP as a distance indicator it is necessary to fit all six clusters with the same plane. We find the best fitting plane for the supercluster is given by:   
\begin{equation} \label{eq:FP_CSC_best_fit}
log\:R_{e}=1.11\:log\:\sigma+0.339\:\langle\mu\rangle_{e}-8.853
\end{equation} 
Direct fits tend to be shallower in $log\:\sigma$ than the inverse or orthogonal fits more typically reported in the literature, but our global best fit is consistent with the direct fit coefficient reported by \cite{Bernardi03c} for their SDSS magnitude limited sample. We find a slightly higher value for $b$ than is reported by \cite{Bernardi03c}, however ours is more consistent with those reported elsewhere in the literature \citep{Jorgensen96}. 

\begin{figure}
\begin{center}
\includegraphics{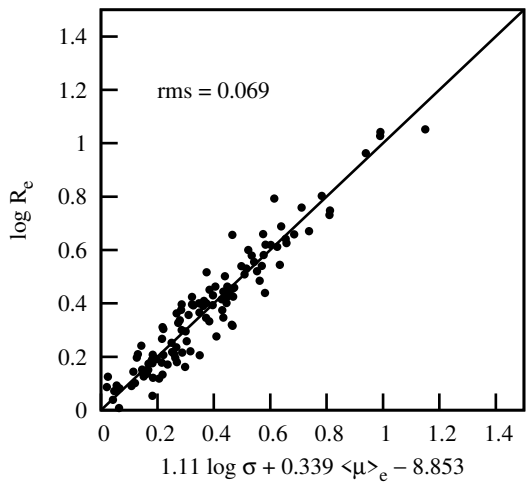}
\caption{Best fitting FP for all 122 galaxies in the sample, with rms dispersion. $R_{e}$ is measured in physical units of $h^{-1}\:kpc$.}
\label{fig:FP}
\end{center}
\end{figure}

\section{Distances and Peculiar Velocities}
\label{sec:dist}
The FP does not directly measure cluster distances; instead the zero-point offsets can be used to find relative distances between clusters. If all clusters in the sample are fitted using the same plane (equation \ref{eq:FP_CSC_best_fit} for our sample) then equation \eqref{eq:FP} can be shown to give:

\begin{equation} \label{eq:distance}
\dfrac{D_{2}}{D_{1}}=10^{c_{1}-c_{2}}
\end{equation}
If the distance to one cluster is known then it can be used to calibrate the relation to give an absolute distance to the queried cluster, where $D_{1}$ is the calibrating distance, $D_{2}$ is the queried distance, both in Mpc, and $c_{1}$ and $c_{2}$ are the zero point offsets for the calibrating and queried clusters, respectively. Calibration is, then, essential in obtaining accurate distance estimates. The magnitude limits on the SDSS mean that it is not possible to calibrate the relations using a nearby cluster with a known distance, such as the Coma cluster, as is often done. Instead we start with the assumption that the clusters are, on average, at rest with respect to the CMB. In other words, we assume no peculiar motion for the supercluster centroid. We then iteratively reduce (using the biweight method) both the cluster redshifts and the cluster offsets to the supercluster centroid, thereby using the supercluster centroid to calibrate the relation. Table \ref{tab:coeffs} shows the offset for the supercluster. The redshift of the supercluster centroid is shown in Table \ref{tab:posdata}.

The individual cluster offsets are shown in Table \ref{tab:coeffs}. In order to avoid any assumptions about the underlying distribution of galaxies in a given cluster, the offsets are found by fitting the data with the plane shown in equation \eqref{eq:FP_CSC_best_fit} and taking a biweight average \citep{Beers} of the individual galaxy offsets. The biweight location estimator provides a more robust estimate of central location than the mean or median and is therefore preferred when the underlying distribution is unknown. Likewise, the biweight scale estimator should be preferred to the rms dispersion for the same reason, but we find that the results from the two methods do not differ significantly, so we follow convention and quote the rms dispersion in the fit.

Figure \ref{fig:FP} shows the best fitting FP for all 122 galaxies in the supercluster sample. All galaxies have been corrected to the same inertial reference frame, with $R_{e}$ in physical units of $h^{-1}\:kpc$. The dispersion in the relation is $0.069$, equivalent to $\sim16$ per cent error in distance to a single galaxy. Table \ref{tab:coeffs} shows the measured cluster distances as FP redshifts. The distances have been calculated using equation \eqref{eq:distance} and converted to redshifts using the approximation \citep{Peebles93}:
\begin{equation}
D=\dfrac{c z}{H_{0}} \left(1- \dfrac{(1+q_{0})z}{2}\right) 
\end{equation}
Error estimates for the cluster distances, based on the dispersion in Figure \ref{fig:FP} and the sample size given in Table \ref{tab:posdata}, are shown in column (6) of Table \ref{tab:coeffs}. Even with the variation in sample size the errors in the distance estimates do not exceed 5 per cent for any cluster, but it is worth reiterating that these error estimates are in the \textit{relative} cluster distances, i.e. they do not account for any error in the calibration.
Peculiar velocities are measured for each cluster based on the difference between photometric and spectroscopic redshift \citep{Danese80}:
\begin{equation}\label{eq:V_pec}
V_{pec}=c \left( \dfrac{ z_{spec}-z_{FP}}{1+z_{FP} }\right)  
\end{equation}
The cluster peculiar velocities are shown in column (7) of Table \ref{tab:coeffs}, where a negative velocity indicates motion towards us, and a positive velocity indicates motion away from us. 

\section{Systematic Errors and Biases}
\label{sec:systematics}
As has been described (\S\ref{sec:fitting}), a number of factors can act to bias the FP coefficients. All of these biases should be a significant concern if one wishes to measure the ``true" FP. However, if the FP is to be used as a secondary distance indicator then these concerns can be largely ignored, provided the biases affect the entire sample similarly. Therefore, when considering sources of error in our distance estimates, we are concerned only with systematic differences among the cluster samples, and how those differences may bias our results and contribute to the error in the FP distances.

All the data are drawn from the same source, and corrections are applied consistently (see \S\ref{sec:data}), so no biases emerge as a result of different data reduction techniques or observational constraints. The range in redshift for the full cluster sample is comparatively narrow, so evolutionary effects can be neglected \citep{Bernardi03c} as can selection effects, as these will also uniformly affect the sample. Care has been taken to avoid contamination by disc galaxies (\S\ref{sec:data}), though it is worth noting that, even if this method has not been completely successful, more recent work suggests that the systematic differences between the best fitting FP for E and S0 galaxies are small enough to be negligible \citep{Donofrio08}. Our velocity dispersion cut of $\sigma\:=\:100\:km\:s^{-1}$ avoids the systematically lower $a$ values seen when low velocity dispersion galaxies are included in the sample (e.g. \cite{Gargiulo09}). All the cluster samples cover similar ranges in $log\: R_{e}$, $\langle\mu\rangle_{e}$, $log\:\sigma$, and absolute magnitude ($M_{r}$), so the same portion of the FP is being sampled (though not uniformly so) in all cases. The rms dispersion in our fit is low, but consistent with fits made using only cluster galaxies and where care has been taken to make sure the sample is homogeneous and as un-biased as possible \citep{Fraix10}. Taking into account the mean observational errors (estimated by the SDSS pipeline) in the measured quantities in the FP, we estimate that measurement error contributes $0.045$ to the dispersion in the fit. This is consistent with the presence of an additional ``intrinsic" component in the scatter, which is generally observed in the FP.  

It is clear that both the scatter and the best fitting coefficients for any given sample are strongly dependent on the galaxies in the sample. It is also clear from the literature that robust fits can only be made to samples that are homologous and ``complete in luminosity, volume, cluster area coverage and stellar kinematics" \citep{Donofrio08}. While we have taken care to make our sample as homogeneous and free of bias as possible, it is obvious that our sample is too small to meet these criteria. Figure \ref{fig:sampling}, which shows the distribution of the FP sample on the sky, also shows the relatively sparse sampling of the clusters. Consequently we would like to understand the possible impact of any intrinsic differences between cluster populations, effects due to non-homology, as well as systematic variations due to sparse and variable sampling, on our results.

Perhaps the most significant assumption implicit in our use of the FP is that all ETGs are drawn from the same underlying distribution. This assumption is supported by \cite{Bernardi03b}, who showed that the joint distribution in $log\:R_{e}$, $log\:\sigma$ and $M_{r}$ for ETGs in the SDSS sample is well described by a trivariate Gaussian, but none the less we would like to test for any systematic differences among the cluster populations that may arise from poor sampling or selection biases. Table \ref{tab:systematics} shows the mean values of the FP observables for each cluster. Variations in these values are to be expected, given the size of the cluster samples, regardless of whether or not the samples are drawn from a common distribution, so it is necessary to determine if these cluster-to-cluster differences are statistically significant. We perform an Anderson--Darling k-sample test on all the galaxy properties ($log\:R_{e}$, $\langle\mu\rangle_{e}$, $log\:\sigma$, $M_{r}$ and the offsets, $c$), as well as subsets of those properties. The Anderson-Darling test is chosen instead of the more generally used Kolmogorov--Smirnoff test as it is more sensitive to variations between distributions \citep{Hou09}. It should be noted, however, that the results presented here do not change if the Kolmogorov--Smirnoff test is used in place of the Anderson--Darling test. The k-sample test is an extension that allows all the clusters to be considered at once, instead of as pairs, as well as testing both multivariate and univariate distributions (available in R \citep{R} under the \textit{adk} package \citep{adk}). Comparing all six samples in each observable independently, we find that the ``null hypothesis" of a single underlying distribution for each parameter is supported in all cases except for $\langle\mu\rangle_{e}$, where it is rejected at a high level of significance ($P=0.008$). Likewise we find that when the joint distribution of $log\:R_{e}$, $log\:\sigma$ and $\langle\mu\rangle_{e}$ is considered, the null hypothesis is also rejected ($P\:=\:0.04$). These results suggest some intrinsic differences among the cluster samples, which we investigate further by performing two-sample tests (both Anderson--Darling and Kolmogorov--Smirnoff) on each of the 15 possible pairs of clusters, for each parameter. Interestingly we find that the null hypothesis is rejected only in cases involving A2079, and when considering only $\langle\mu\rangle_{e}$, it is rejected in almost \textit{every} case involving A2079 (and it is rejected in absolutely every case when the Kolmogorov--Smirnoff test is used). Excluding A2079 from the k-sample test we find that the null hypothesis is accepted in every case, with $P=0.50$ for the joint distribution of $log\:R_{e}$, $log\:\sigma$, and $\langle\mu\rangle_{e}$, and $P=0.57$ for the distribution of $\langle\mu\rangle_{e}$. 

\begin{table*}
\begin{minipage}{150mm}
\caption{Mean values of the FP observables, as well as absolute magnitude, for each of the clusters. Columns (6) and (7) give the FP distances and resulting peculiar velocities when A2079 is excluded from the fit. The errors in the distances are the same as those shown in Table \ref{tab:coeffs}. The last two columns show the FP distances and peculiar velocities for the clipped sample, with A2079 included.}
\label{tab:systematics}
\begin{center}
\begin{tabular}{ccccccccc}
\hline 
Cluster & $\langle log\:R_{e}\rangle$ & $\langle\mu_{e}\rangle$ & $\langle log\:\sigma\rangle$ & $\langle M_{r}\rangle$ & $z_{FP}$ & $V_{pec}\:(km\:s^{-1})$ & $z_{c_{FP}}$ & $V_{pec} (km\:s^{-1})$ \\ 
\hline 
A2092 & 0.375 & 19.90 & 2.239 & -20.72 & 0.0686 & -516 & 0.0717 &-1490 \\ 
A2089 & 0.293 & 19.77 & 2.209 & -20.63 & 0.0711 & 835 & 0.0705 & 891\\ 
A2079 & 0.246 & 19.41 & 2.245 & -20.74 & -- & -- & 0.0654 & 140\\
A2067 & 0.503 & 20.16 & 2.270 & -20.23 & 0.0728 & 448 & 0.0720 & 560\\ 
A2065 & 0.382 & 19.97 & 2.220 & -20.74 & 0.0721 & 8 & 0.0732 & -411\\ 
A2061 & 0.359 & 20.00 & 2.202 & -20.86 & 0.0726 & 1850 & 0.0712 & 2130 \\ 
\hline 
\end{tabular}
\end{center} 
\end{minipage}
\end{table*}

These results suggest that the sample for A2079 may systematically deviate from the other cluster samples (which these tests indicate are drawn from a common distribution). Moreover, when looking at the scatter in the best fitting FP for the individual clusters (shown in Table \ref{tab:coeffs}) the scatter in the fit for A2079 is markedly higher than for the other clusters, with rms $=0.081$. It would be reasonable to assume that this systematic difference may bias all the distance estimates, instead of only that of A2079. However, when we perform the FP analysis excluding A2079 we find that very little changes in the results. Table \ref{tab:systematics} shows the FP distances and peculiar velocities for the $5$ clusters (columns (6) and (7)), with the best fitting plane given by:

\begin{equation} \label{eq:FP_no79_best_fit}
log\:R_{e}=1.13\:log\:\sigma+0.352\:\langle\mu\rangle_{e}-9.163
\end{equation}

The values of both $a$ and $b$ are consistent with those of the original fit (equation \ref{eq:FP_CSC_best_fit}), and we see a systematic reduction in the offsets of $\sim0.3$. The intrinsic dispersion in this fit is $0.067$, which is negligibly different from that in Figure \ref{fig:FP}, and the distances and peculiar velocities are consistent with those in Table \ref{tab:coeffs}. This suggests that while the A2079 sample does, on average, have systematically higher mean surface brightness, the inclusion of this sample in the global fit does not significantly bias the distance estimates for the other clusters. 

Non-homology is often considered as a cause of the tilt of the FP \citep{Gargiulo09} from the virial plane, so it is worth considering how this might affect our sample and if we might expect to see any systematic differences between the clusters. In their study of 56 low-redshift clusters \cite{Fraix10} took a novel approach to investigating the role of non-homology in the intrinsic dispersion in the FP, using astrocladistics to identify independent homologous (defined as similar ``due to the same class of progenitor") subgroups in a sample of 699 ETGs from the SMAC catalog. The method of astrocladistics was chosen because it identifies homologous groups in a sample without making any a priori assumptions about which properties are most closely linked with homology (see reference for details of the method), and the sample was chosen because great care had been taken to make it homogeneous and as free as possible of systematic biases. As with our work the best fitting FP is found using a direct fit in $log\:R_{e}$ and the cluster samples are similar in size to our own. The best fitting FP that they find for their full sample is highly consistent with our own, and the scatter in the relation is only marginally higher in our sample than in theirs ($0.069$ compared with $0.065$). Seven homologous groups are identified within their sample, and they find that all clusters contain galaxies from all groups, concluding that, when considering non-homology as a source of bias or scatter, intra-cluster differences are far greater than inter-cluster differences. We conclude that non-homology is unlikely to be a significant source of systematic bias in our results, and the scatter in the FP likely gives a good approximation of the error in distance estimates due to the effects of non-homology.

Investigations of the intrinsic scatter in the FP have demonstrated a small but significant environmental dependence of the coefficients \citep{Donofrio08,Labarb10b,Hyde09}. This trend is, of course, strongest when comparing field galaxies with cluster galaxies, but both \cite{Donofrio08} and \cite{Labarb10b} find a smooth trend with local density within clusters, and \cite{Labarb10b} suggest that this could cause a systematic variation in the offsets of $\sim0.02$ for every decade change in local density. It is clear from Figure \ref{fig:sampling} that the clusters are not uniformly sampled, and any significant variation in mean local density between the cluster samples may be a source of systematic bias in our results. It is difficult to assess the contribution of this possible systematic error to the total error in the distance estimates, but Figure \ref{fig:sampling} shows that several of the clusters contain apparent outliers, galaxies that are right at the limit of our defined cluster radius. Clipping those data from the sample (7 points in total: 2 each from A2092 and A2089, and 1 each from A2067, 65 and 61) and fitting the FP once again causes a significant increase in $a$ ($a=1.24$). However, little of substance changes in the distance determinations, except in the case of A2079, which is now placed effectively at rest. It is worth noting that when we fit the FP to this sample and perform an iterative $3\:\sigma$ clipping to reject any outliers, an additional two data points are eliminated from A2079 and one more is eliminated from A2092 ($3\:\sigma$ clipping does not reject any data points in the full sample). This clipping does not meaningfully change the results for A2079, though it contributes to a slight increase in the estimated distance of A2092.

Table \ref{tab:systematics} shows the results for the clipped sample (columns (8) and (9)). The dispersion in this sample is reduced to $0.063$, which has the net effect of keeping the statistical error estimates the same as those for the full sample, and the distances are consistent with those determined from the full sample. The results do not change when A2079 is excluded from the analysis. The sensitivity of the distance measurement of A2079 to changes in the overall sample further suggests that our results for this cluster must be viewed with caution, and the corresponding error estimate may well be underestimated. While this analysis does not directly address the possible systematic differences in local environment between the cluster samples, it does demonstrate a certain robustness to small changes based on distance from the cluster center, which can be viewed as a proxy for local density \citep{Labarb10b} (except in the case of A2079, of course).

\section{The Kormendy Relation}
\label{sec:KR}
There exist several scaling relations between observables for ETGs (\S\ref{sec:SDI}) that can be used as secondary distance indicators. These relations are two dimensional projections of the FP and as such are generally inferior in both accuracy and precision (with the exception of $D_{n}-\sigma$, which is comparable to the FP in that it effectively combines effective radius and central surface brightness in the parameter $D_{n}$ \citep{Djorgovski87}). However these relations can act as useful alternatives to the FP when the availability of certain types of data is limited. The Kormendy Relation (KR) \citep{Kormendy77} is the projection of the FP relating effective radius and central surface brightness:
\begin{equation} \label{eq:KR}
log\:r_{e} = \alpha\:\langle\mu\rangle_{e}+\beta
\end{equation}
This can be used as a secondary distance indicator in much the same way as the FP, with the zero point offsets, $\beta$, giving a direct indication of distance, once suitably calibrated. Scatter in the KR is considerably higher than in the FP, equivalent to $\sim30$ per cent error in distance to a single galaxy, and the results are potentially more susceptible to biases introduced by correlated errors, as well as systematic differences between cluster samples, which might be offset in the FP by the additional constraint on velocity dispersion (see \S\ref{subsec:KRdist}). However, the KR retains the significant advantage over the FP of requiring only photometric data, which is both more abundantly available and significantly easier to collect than the quality spectroscopic data needed to measure velocity dispersions. 

The data set for the KR analysis is slightly larger than for the FP since we no longer have a constraint on velocity dispersion, and in almost all cases there are more galaxies with high quality photometry than with the high quality spectra necessary to measure velocity dispersions. The data have been corrected in exactly the same way as described in \S\ref{sec:data} and Table \ref{tab:KRcoeffs} gives the sample sizes for each cluster.

\subsection{Fitting the KR}
As with the FP, a direct fit requires determination of $\alpha$ such that
\begin{equation}\label{eq:KR_residual}
\sum\limits_{i} \left(log\:r_{e_{i}}-\alpha\:\langle\mu\rangle_{e_{i}}-\beta_{i}\right)^{2}
\end{equation}
 summed over all $N$ galaxies in the cluster, is minimized. The best fitting coefficients are given by
\begin{eqnarray}
& \alpha=\dfrac{\rho_{\mu R}^{2}}{\rho_{\mu \mu}^{2}}\nonumber\\
& \beta_{i}=log\:r_{e_{i}}-\alpha\:\langle\mu\rangle_{e_{i}}
\end{eqnarray} 
Corrections are applied as in \S\ref{sec:fitting} to achieve an unbiased estimate of the slope. The offsets $\beta_{i}$ are calculated for each galaxy and then averaged, as for the FP, to get cluster offsets. Table \ref{tab:KRcoeffs} shows the best fitting KR coefficients for each cluster. Similarly to the results for the FP we see significant variation in $\alpha$ from cluster to cluster, with the best fitting relation for the supercluster given by
\begin{equation} \label{eq:KR_CSC_best_fit}
log\:R_{e}=0.281\:\langle\mu\rangle_{e}-5.223
\end{equation}
The coefficients of the KR are generally less sensitive to fitting method than those of the FP (though this can be seen as a similarity with the FP in that $a$ is generally far more sensitive to fitting method than $b$), and this fit is completely consistent with that of \cite{Labarb10b}. 
\subsection{Distances and Peculiar Velocities}
\label{subsec:KRdist}
\begin{figure}
\begin{center}
\includegraphics{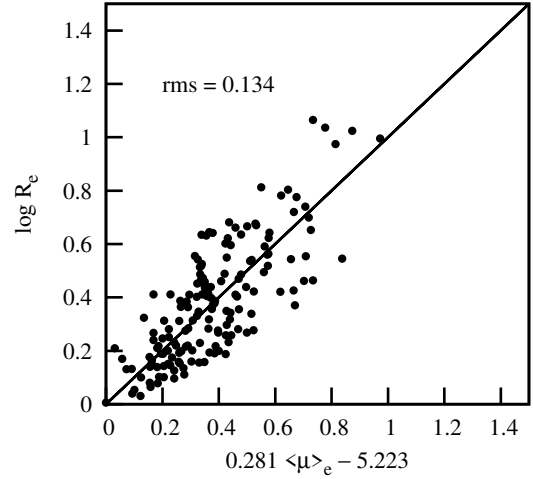}
\caption{Best fitting KR for all 166 galaxies in the sample, with rms dispersion. $R_{e}$ is measured in physical units of $h^{-1}\:kpc$.}
\label{fig:KR}
\end{center}
\end{figure}
\begin{table*}
\begin{minipage}{150mm}
\caption{Best fitting KR coefficients for each of the clusters, with associated errors (calculated via a bootstrapping procedure), are shown in columns (3) and (4), with the last row giving the best fitting coefficients and associated errors for the supercluster. Columns (5) and (6) give the cluster distances determined from the KR ($z_{KR}$) and associated error (determined from the rms dispersion in the global fit), respectively, and column (7) gives the resulting peculiar velocities.}
\label{tab:KRcoeffs}
\begin{center}
\begin{tabular}{ccccccc}
\hline 
Cluster & $N_{KR}$ & $\alpha$ & $\beta$ & $z_{KR}$ & error ($\%$)& $V_{pec}\: (km\:s^{-1})$ \\
\hline 
A2092 & 21 & $0.315\pm 0.049$ & $-5.250\pm 0.025$ & $0.0712$ & 7 & $-1354$ \\ 
A2089 & 25 & $0.296\pm 0.044$ & $-5.225\pm 0.032$ & $0.0741$ & 6 & $-116$ \\ 
A2079 & 21 & $0.271\pm 0.058$ & $-5.218\pm 0.033$ & $0.0656$ & 7 & $84$ \\ 
A2067 & 13 & $0.287\pm 0.082$ & $-5.186\pm 0.058$ & $0.0678$ & 9 & $1740$ \\ 
A2065 & 53 & $0.248\pm 0.028$ & $-5.240\pm 0.017$ & $0.0748$ & 4 & $-856$ \\ 
A2061 & 33 & $0.270\pm 0.048$ & $-5.211\pm 0.033$ & $0.0762$ & 5 & $728$ \\
CSC & 166 & $0.281\pm 0.01$ & $-5.223\pm 0.011$ &  $0.0717$ & -- &--\\
\hline 
\end{tabular}
\end{center}
\end{minipage}
\end{table*}
The cluster offsets $\beta$ are shown in Table \ref{tab:KRcoeffs}. As with the FP, these offsets were found by taking the biweight average of the galaxy offsets for each cluster, where the data had been fitted using the slope from equation  \eqref{eq:KR_CSC_best_fit}. Figure \ref{fig:KR} shows the best fitting KR for all 166 galaxies in the supercluster sample. All galaxies have been corrected to the same inertial reference frame, with $R_{e}$ in $h^{-1}\:kpc$. The dispersion in this relation is $0.134$, equivalent to $\sim31$ per cent error in distance to a single galaxy. The cluster distances from this relation are shown in column (5) of Table \ref{tab:KRcoeffs} as KR redshifts, calculated as described for the FP, with errors given in column (6). Though the sample sizes are a little larger for the KR than for the FP, the larger dispersion in the relation results in larger errors overall, with mean accuracy in the distance determinations of $\sim6$ per cent. Peculiar velocities, calculated using equation \eqref{eq:V_pec}, are shown in column (7) of Table \ref{tab:KRcoeffs}. The sign of the velocity indicates direction of motion as described in \S\ref{sec:dist}. 

Comparing the results in Tables \ref{tab:coeffs} and \ref{tab:KRcoeffs} it is clear that the two methods give different results, though in all cases they are consistent within the errors. Excepting A2067, we find generally that the KR distances are systematically larger than the FP distances, by $\sim0.002-0.004$. In contrast we find that the KR distance to A2067 is quite a lot smaller than the FP distance, by $\sim0.005$. Comparing results from the FP and the KR for the pair A2061 and A2067 is particularly interesting because, not only do those results display the largest variability of all the clusters, but the clusters also ``switch" positions (i.e. which is foreground and which is background) depending on which method is used. This might imply that the two samples are contaminating each other, and that our methods for separating the structures are flawed. However, we find this unlikely since the redshift range for both clusters is very narrow, and the redshifts that we find agree well with those previously published, suggesting that we are likely only using core galaxies in our analysis. 

We suggest that sparse sampling is the most likely cause of the systematic differences between the FP and the KR. The analysis of the environmental dependence of the FP coefficients of \cite{Labarb10b} shows clearly that, for high mass groups and clusters, $b$ (and thus also $\alpha$) is increasingly (with cluster mass) dependent on local density, while the dependence of $a$ not only weakens but shows a trend which opposes that of $b$. This suggests that, in the high cluster mass regime, the FP offsets should be more robust to variations in mean local density than the KR offsets. Cluster-to-cluster variations in mean local density might well be expected in our sample (see discussion in \ref{sec:systematics}), and we would consequently expect the KR to be more sensitive than the FP to those variations. This may be particularly apparent with A2061 and A2067 because, while the clusters are similarly massive and very close together, A2061 is substantially better sampled than A2067. 

\section{Discussion}
\label{sec:disc}
The results from the FP analysis are summarized in Table \ref{tab:coeffs}. Significant peculiar velocities are indicated for all clusters except A2065. In the case of A2065 the FP distance is consistent with the spectroscopic redshift, so the relatively small peculiar velocity of $-229\:km\:s^{-1}$ that we find for this cluster is consistent with no significant peculiar motion relative to the supercluster centroid. 
By far the most significant peculiar velocity is indicated for A2061. Given the proximity of A2067 and A2061 on the sky (Figure \ref{fig:sampling}), as well as the fact that along with A2065 they are the largest mass concentrations in the supercluster (Table \ref{tab:posdata} shows the cluster masses), it is perhaps most interesting to look at these two clusters as a possible bound pair. The spectroscopic redshifts place them at some distance from each other (in redhsift space), but the results of the FP suggest that they are substantially closer in real space. \cite{Marini04} investigated the possibility that this was a merging pair using X-ray observations of the two clusters. They found evidence that A2061 is significantly elongated towards A2067, but concluded that the line-of-sight distance between them (based on spectroscopic redshifts) was too great for the elongation to have been caused by gravitational interaction with A2067. Our results suggest that not only is the physical separation along the line of sight actually far smaller than the separation in redshift, but the difference in the spectroscopic redshifts ($cz\approx1440\:km\:s^{-1}$) is almost identical to the relative peculiar velocity of the two clusters ($\sim1485\:km\:s^{-1}$). This is consistent with A2067 and A2061 being a very close bound pair, where the difference in spectroscopic redshift is due to significant gravitational interaction, and together they constitute the largest mass concentration in the CSC. If this is the case, and if the core of the CSC is a single bound structure, then we might expect A2065, given its mass and proximity to this pair both in redshift and on the sky, to be bound to it. The absence of any significant peculiar velocity for A2065 indicated by our analysis suggests that it is dynamically isolated from A2067 and A2061, which leads us to conclude that the core of the CSC is not a single bound structure.

The peculiar velocities for both A2089 and A2092 are large enough to suggest gravitational interactions, though larger samples are certainly necessary to confirm this. These are the lowest mass clusters in the sample, and consequently we would expect that, if they are part of the bound structure, they would be bound to the largest nearby mass concentrations. The peculiar velocities we have determined are certainly consistent with this picture. The position of A2089, both in redshift and on the sky, suggests that it is likely bound to A2065 and thus probably has a significant tangential component to its peculiar velocity. The position of A2092 suggests that it is likely bound to the pair A2061 and A2067. 

The peculiar velocity determined for A2079 is significant but, given the results described in \S\ref{sec:systematics}, we must view it cautiously. It is likely that the systematic deviation of the A2079 population from the rest of the sample biases the results such that we refrain from making any definitive statements about the motion of A2079, except to say that the cluster mass, as well as its position on the sky, does not exclude the possibility that it is bound to A2065 and A2089. The peculiar velocity that we find certainly supports this notion and, if it is accurate, it further indicates that this region of the CSC has reached turnaround and is in collapse. More complete observations of the cluster are needed in order to properly investigate this possibility.

In general our results support the conclusions of previous authors \citep{Postman88,small3,Kopylova98} that the CSC is a bound structure. However in contrast to the results of \cite{Kopylova98}, who identified a single bound region consisting of A2065, 2067, 2061, 2089 and 2092, we find evidence for two bound regions in the core of the CSC. It is also worth noting that our derived peculiar velocities differ significantly from those of \cite{Kopylova98}, which is not surprising given the problems inherent in using the KR with sparsely sampled clusters. Investigations by \cite{Bahcall94} suggested that, in general, cluster peculiar velocities in dense superclusters could range from $600\:km\:s^{-1}$ to as high as $2000\:km\:s^{-1}$. In comparing our derived peculiar velocities with those of \cite{Kopylova98} we find that in general our peculiar velocities are smaller, placing them well within the ranges given by both \cite{Postman88} ($\leq\:2200\:km\:s^{-1}$) and \cite{Bahcall94}.

That we observe peculiar velocities at all is interesting in light of the recent work by \cite{Pearson13}, who performed N-body simulations of the CSC using the four clusters in the densest part of the core (excluding A2079 and A2092). The clusters were modeled as point particles, with masses based on velocity dispersions gleaned from the literature, and Monte Carlo simulations were run using initial conditions based on currently available data and error estimates. Their results indicate that there is insufficient mass in the clusters to bind any part of the CSC, and even the close pair A2061 and A2067 is predicted as unlikely to be bound by their simulations. This suggests that a dominant mass component outside the clusters is required to cause the peculiar velocities that we see, and it is not clear that inter-cluster galaxies could contribute sufficient mass to explain the discrepancy.\footnote{\cite{Proust06} found that inter-cluster galaxies in the SSC might contribute up to twice as much mass to the supercluster as the cluster galaxies. This may explain part of the discrepancy, but is unlikely to explain the peculiar velocities that we observe (D.~W.~Pearson, private communication).} Further investigation is required to understand and resolve this inconsistency, but one possible explanation is a significant inter-cluster dark matter component. The presence of a dominant inter-cluster dark matter component has been suggested in the SSC \citep{Quintana95}, while more generally it has been suggested that dark matter filaments may exist between close rich clusters \citep{Gray02}, and this would certainly explain our observations, suggesting that gravitationally bound superclusters may be an excellent environment in which to search for inter-cluster dark matter. 

\section{Conclusions}
\label{sec:conc}
We have performed a dynamical analysis of the Corona Borealis Supercluster, assessing the peculiar motions of six clusters in the densest region of the supercluster. Using data from the SDSS and the Fundamental Plane relation, we have made accurate estimates of the relative distances to these clusters, independent of redshift, and have used them to derive peculiar velocities for these clusters. We find that four of the six clusters have significant line-of-sight peculiar velocities, providing observational evidence that the CSC contains two bound regions which have reached turnaround and are in collapse. This is only the second bound supercluster to have been identified in the universe, along with the SSC at $z\approx 0.05$. In comparing with recent simulation work we find that the observed peculiar velocities cannot be explained by the masses of the clusters alone, suggesting significant contribution to the mass of the supercluster from the inter-cluster region. We suggest that this may be due in part to an inter-cluster dark matter component, though extensive further investigation is required to confirm such a possibility. An alternative analysis using the Kormendy Relation is also presented, with the intention of providing a baseline for comparison with future work. In all cases the results from both the FP and the KR are consistent within the errors. However, a systematic shift between the distances derived by the two methods of $0.002-0.004$ is apparent, with the KR distances being consistently higher than the FP distances in all but one case. We conclude that the differences between the two methods result from sparse sampling of the clusters, to which the KR is likely more sensitive than the FP. This suggests that completeness in cluster area coverage, whilst important for both methods, is essential if the KR is to be used to obtain accurate redshift independent distances. 
\section*{Acknowledgements}
We thank D.W. Pearson for helpful discussion, and insight regarding simulation work. This research was partially supported by grants from the University of Maine, in conjunction with the Maine Space Grant Consortium (MSGC). This research has made use of the NASA/IPAC Extragalactic Database (NED) which is operated by the Jet Propulsion Laboratory, California Institute of Technology, under contract with the National Aeronautics and Space Administration.
\bibliography{ref} 

\begin{thebibliography}{}

\bibitem[\protect\citeauthoryear{{Abazajian}, {Adelman-McCarthy},
  {Ag{\"u}eros}, {Allam}, {Anderson}, {Anderson}, {Annis}, {Bahcall} \& {et
  al.}}{{Abazajian} et~al.}{2004}]{SDSS2}
{Abazajian} K.,  {Adelman-McCarthy} J.~K.,  {Ag{\"u}eros} M.~A.,  {Allam}
  S.~S.,  {Anderson} K.,  {Anderson} S.~F.,  {Annis} J.,  {Bahcall} N.~A.,
  {et al.} 2004, AJ, 128, 502

\bibitem[\protect\citeauthoryear{{Abazajian}, {Adelman-McCarthy},
  {Ag{\"u}eros}, {Allam}, {Allende Prieto}, {An}, {Anderson}, {Anderson},
  {Annis}, {Bahcall} \& et al.}{{Abazajian} et~al.}{2009}]{SDSS7}
{Abazajian} K.~N.,  {Adelman-McCarthy} J.~K.,  {Ag{\"u}eros} M.~A.,  {Allam}
  S.~S.,  {Allende Prieto} C.,  {An} D.,  {Anderson} K.~S.~J.,  {Anderson}
  S.~F.,  {Annis} J.,  {Bahcall} N.~A.,    et al. 2009, ApJS, 182, 543

\bibitem[\protect\citeauthoryear{{Abell}}{{Abell}}{1958}]{Abell58}
{Abell} G.~O.,  1958, ApJS, 3, 211

\bibitem[\protect\citeauthoryear{{Akritas} \& {Bershady}}{{Akritas} \&
  {Bershady}}{1996}]{Akritas}
{Akritas} M.~G.,  {Bershady} M.~A.,  1996, ApJ, 470, 706

\bibitem[\protect\citeauthoryear{{Araya-Melo}, {Reisenegger}, {Meza}, {van de
  Weygaert}, {D{\"u}nner} \& {Quintana}}{{Araya-Melo} et~al.}{2009}]{Araya09}
{Araya-Melo} P.~A.,  {Reisenegger} A.,  {Meza} A.,  {van de Weygaert} R.,
  {D{\"u}nner} R.,    {Quintana} H.,  2009, MNRAS, 399, 97

\bibitem[\protect\citeauthoryear{{Bahcall}, {Gramann} \& {Cen}}{{Bahcall}
  et~al.}{1994}]{Bahcall94}
{Bahcall} N.~A.,  {Gramann} M.,    {Cen} R.,  1994, ApJ, 436, 23

\bibitem[\protect\citeauthoryear{{Baldi}, {Bardelli} \& {Zucca}}{{Baldi}
  et~al.}{2001}]{Baldi01}
{Baldi} A.,  {Bardelli} S.,    {Zucca} E.,  2001, MNRAS, 324, 509

\bibitem[\protect\citeauthoryear{{Bardelli}, {Scaramella}, {Vettolani},
  {Zamorani}, {Zucca}, {Collins} \& {MacGillivray}}{{Bardelli}
  et~al.}{1993}]{Bardelli93}
{Bardelli} S.,  {Scaramella} R.,  {Vettolani} G.,  {Zamorani} G.,  {Zucca} E.,
  {Collins} C.~A.,    {MacGillivray} H.~T.,  1993, The Messenger, 71, 34

\bibitem[\protect\citeauthoryear{{Beers}, {Flynn} \& {Gebhardt}}{{Beers}
  et~al.}{1990}]{Beers}
{Beers} T.~C.,  {Flynn} K.,    {Gebhardt} K.,  1990, AJ, 100, 32

\bibitem[\protect\citeauthoryear{{Bernardi}, {Sheth}, {Annis}, {Burles},
  {Eisenstein}, {Finkbeiner}, {Hogg}, {Lupton} \& {et al.}}{{Bernardi}
  et~al.}{2003a}]{Bernardi03a}
{Bernardi} M.,  {Sheth} R.~K.,  {Annis} J.,  {Burles} S.,  {Eisenstein} D.~J.,
  {Finkbeiner} D.~P.,  {Hogg} D.~W.,  {Lupton} R.~H.,    {et al.} 2003a, AJ,
  125, 1817

\bibitem[\protect\citeauthoryear{{Bernardi}, {Sheth}, {Annis}, {Burles},
  {Eisenstein}, {Finkbeiner}, {Hogg}, {Lupton} \& {et al.}}{{Bernardi}
  et~al.}{2003c}]{Bernardi03b}
{Bernardi} M.,  {Sheth} R.~K.,  {Annis} J.,  {Burles} S.,  {Eisenstein} D.~J.,
  {Finkbeiner} D.~P.,  {Hogg} D.~W.,  {Lupton} R.~H.,    {et al.} 2003c, AJ,
  125, 1849

\bibitem[\protect\citeauthoryear{{Bernardi}, {Sheth}, {Annis}, {Burles},
  {Eisenstein}, {Finkbeiner}, {Hogg}, {Lupton} \& {et al.}}{{Bernardi}
  et~al.}{2003b}]{Bernardi03c}
{Bernardi} M.,  {Sheth} R.~K.,  {Annis} J.,  {Burles} S.,  {Eisenstein} D.~J.,
  {Finkbeiner} D.~P.,  {Hogg} D.~W.,  {Lupton} R.~H.,    {et al.} 2003b, AJ,
  125, 1866

\bibitem[\protect\citeauthoryear{{Binney} \& {Merrifield}}{{Binney} \&
  {Merrifield}}{1998}]{BinneyMerri}
{Binney} J.,  {Merrifield} M.,  1998, {Galactic Astronomy}.
Princeton University Press

\bibitem[\protect\citeauthoryear{{Chilingarian}, {Melchior} \&
  {Zolotukhin}}{{Chilingarian} et~al.}{2010}]{kcor}
{Chilingarian} I.~V.,  {Melchior} A.-L.,    {Zolotukhin} I.~Y.,  2010, MNRAS,
  405, 1409

\bibitem[\protect\citeauthoryear{{Danese}, {de Zotti} \& {di Tullio}}{{Danese}
  et~al.}{1980}]{Danese80}
{Danese} L.,  {de Zotti} G.,    {di Tullio} G.,  1980, A\&A, 82, 322

\bibitem[\protect\citeauthoryear{{Djorgovski} \& {Davis}}{{Djorgovski} \&
  {Davis}}{1987}]{Djorgovski87}
{Djorgovski} S.,  {Davis} M.,  1987, ApJ, 313, 59

\bibitem[\protect\citeauthoryear{{D'Onofrio}, {Fasano}, {Varela}, {Bettoni},
  {Moles}, {Kj{\ae}rgaard}, {Pignatelli}, {Poggianti}, {Dressler}, {Cava},
  {Fritz}, {Couch} \& {Omizzolo}}{{D'Onofrio} et~al.}{2008}]{Donofrio08}
{D'Onofrio} M.,  {Fasano} G.,  {Varela} J.,  {Bettoni} D.,  {Moles} M.,
  {Kj{\ae}rgaard} P.,  {Pignatelli} E.,  {Poggianti} B.,  {Dressler} A.,
  {Cava} A.,  {Fritz} J.,  {Couch} W.~J.,    {Omizzolo} A.,  2008, ApJ, 685,
  875

\bibitem[\protect\citeauthoryear{{Dressler}, {Lynden-Bell}, {Burstein},
  {Davies}, {Faber}, {Terlevich} \& {Wegner}}{{Dressler}
  et~al.}{1987}]{Dressler87}
{Dressler} A.,  {Lynden-Bell} D.,  {Burstein} D.,  {Davies} R.~L.,  {Faber}
  S.~M.,  {Terlevich} R.,    {Wegner} G.,  1987, ApJ, 313, 42

\bibitem[\protect\citeauthoryear{{Einasto}, {Joeveer} \& {Saar}}{{Einasto}
  et~al.}{1980}]{Einasto80}
{Einasto} J.,  {Joeveer} M.,    {Saar} E.,  1980, MNRAS, 193, 353

\bibitem[\protect\citeauthoryear{{Ettori}, {Fabian} \& {White}}{{Ettori}
  et~al.}{1997}]{Ettori97}
{Ettori} S.,  {Fabian} A.~C.,    {White} D.~A.,  1997, MNRAS, 289, 787

\bibitem[\protect\citeauthoryear{{Faber} \& {Jackson}}{{Faber} \&
  {Jackson}}{1976}]{FJ76}
{Faber} S.~M.,  {Jackson} R.~E.,  1976, ApJ, 204, 668

\bibitem[\protect\citeauthoryear{{Fraix-Burnet}, {Dugu{\'e}}, {Chattopadhyay},
  {Chattopadhyay} \& {Davoust}}{{Fraix-Burnet} et~al.}{2010}]{Fraix10}
{Fraix-Burnet} D.,  {Dugu{\'e}} M.,  {Chattopadhyay} T.,  {Chattopadhyay}
  A.~K.,    {Davoust} E.,  2010, MNRAS, 407, 2207

\bibitem[\protect\citeauthoryear{{Gargiulo}, {Haines}, {Merluzzi}, {Smith},
  {Barbera}, {Busarello}, {Lucey}, {Mercurio} \& {Capaccioli}}{{Gargiulo}
  et~al.}{2009}]{Gargiulo09}
{Gargiulo} A.,  {Haines} C.~P.,  {Merluzzi} P.,  {Smith} R.~J.,  {Barbera}
  F.~L.,  {Busarello} G.,  {Lucey} J.~R.,  {Mercurio} A.,    {Capaccioli} M.,
  2009, MNRAS, 397, 75

\bibitem[\protect\citeauthoryear{{Gibbons}, {Fruchter} \& {Bothun}}{{Gibbons}
  et~al.}{2001}]{Gibbons01}
{Gibbons} R.~A.,  {Fruchter} A.~S.,    {Bothun} G.~D.,  2001, AJ, 121, 649

\bibitem[\protect\citeauthoryear{{Gray}, {Taylor}, {Meisenheimer}, {Dye},
  {Wolf} \& {Thommes}}{{Gray} et~al.}{2002}]{Gray02}
{Gray} M.~E.,  {Taylor} A.~N.,  {Meisenheimer} K.,  {Dye} S.,  {Wolf} C.,
  {Thommes} E.,  2002, ApJ, 568, 141

\bibitem[\protect\citeauthoryear{{Hou}, {Parker}, {Harris} \& {Wilman}}{{Hou}
  et~al.}{2009}]{Hou09}
{Hou} A.,  {Parker} L.~C.,  {Harris} W.~E.,    {Wilman} D.~J.,  2009, ApJ, 702,
  1199

\bibitem[\protect\citeauthoryear{{Hyde} \& {Bernardi}}{{Hyde} \&
  {Bernardi}}{2009}]{Hyde09}
{Hyde} J.~B.,  {Bernardi} M.,  2009, MNRAS, 394, 1978

\bibitem[\protect\citeauthoryear{{J{\o}rgensen}, {Franx} \&
  {Kj{\ae}rgaard}}{{J{\o}rgensen} et~al.}{1995}]{Jorgensen95}
{J{\o}rgensen} I.,  {Franx} M.,    {Kj{\ae}rgaard} P.,  1995, MNRAS, 276, 1341

\bibitem[\protect\citeauthoryear{{J{\o}rgensen}, {Franx} \&
  {Kj{\ae}rgaard}}{{J{\o}rgensen} et~al.}{1996}]{Jorgensen96}
{J{\o}rgensen} I.,  {Franx} M.,    {Kj{\ae}rgaard} P.,  1996, MNRAS, 280, 167

\bibitem[\protect\citeauthoryear{{Kopylova} \& {Kopylov}}{{Kopylova} \&
  {Kopylov}}{1998}]{Kopylova98}
{Kopylova} F.~G.,  {Kopylov} A.~I.,  1998, Astronomy Letters, 24, 491

\bibitem[\protect\citeauthoryear{{Kormendy}}{{Kormendy}}{1977}]{Kormendy77}
{Kormendy} J.,  1977, ApJ, 218, 333

\bibitem[\protect\citeauthoryear{{La Barbera}, {de Carvalho}, {de La Rosa} \&
  {Lopes}}{{La Barbera} et~al.}{2010}]{Labarb10b}
{La Barbera} F.,  {de Carvalho} R.~R.,  {de La Rosa} I.~G.,    {Lopes}
  P.~A.~A.,  2010, MNRAS, 408, 1335

\bibitem[\protect\citeauthoryear{{La Barbera}, {Lopes}, {de Carvalho}, {de La
  Rosa} \& {Berlind}}{{La Barbera} et~al.}{2010}]{Labarb10}
{La Barbera} F.,  {Lopes} P.~A.~A.,  {de Carvalho} R.~R.,  {de La Rosa} I.~G.,
    {Berlind} A.~A.,  2010, MNRAS, 408, 1361

\bibitem[\protect\citeauthoryear{{Marini}, {Bardelli}, {Zucca}, {De Grandi},
  {Cappi}, {Ettori}, {Moscardini}, {Tormen} \& {Diaferio}}{{Marini}
  et~al.}{2004}]{Marini04}
{Marini} F.,  {Bardelli} S.,  {Zucca} E.,  {De Grandi} S.,  {Cappi} A.,
  {Ettori} S.,  {Moscardini} L.,  {Tormen} G.,    {Diaferio} A.,  2004, MNRAS,
  353, 1219

\bibitem[\protect\citeauthoryear{{Pearson} \& {Batuski}}{{Pearson} \&
  {Batuski}}{2013}]{Pearson13}
{Pearson} D.~W.,  {Batuski} D.~J.,  2013, MNRAS, doi:10.1093/mnras/stt1614

\bibitem[\protect\citeauthoryear{{Peebles}}{{Peebles}}{1993}]{Peebles93}
{Peebles} P.~J.~E.,  1993, {Principles of Physical Cosmology}.
Princeton University Press

\bibitem[\protect\citeauthoryear{{Postman}, {Geller} \& {Huchra}}{{Postman}
  et~al.}{1988}]{Postman88}
{Postman} M.,  {Geller} M.~J.,    {Huchra} J.~P.,  1988, AJ, 95, 267

\bibitem[\protect\citeauthoryear{{Proust}, {Quintana}, {Carrasco},
  {Reisenegger}, {Slezak}, {Muriel}, {D{\"u}nner}, {Sodr{\'e}} Jr.,
  {Drinkwater}, {Parker} \& {Ragone}}{{Proust} et~al.}{2006}]{Proust06}
{Proust} D.,  {Quintana} H.,  {Carrasco} E.~R.,  {Reisenegger} A.,  {Slezak}
  E.,  {Muriel} H.,  {D{\"u}nner} R.,  {Sodr{\'e}} Jr. L.,  {Drinkwater} M.~J.,
   {Parker} Q.~A.,    {Ragone} C.~J.,  2006, A\&A, 447, 133

\bibitem[\protect\citeauthoryear{{Quintana}, {Ramirez}, {Melnick},
  {Raychaudhury} \& {Slezak}}{{Quintana} et~al.}{1995}]{Quintana95}
{Quintana} H.,  {Ramirez} A.,  {Melnick} J.,  {Raychaudhury} S.,    {Slezak}
  E.,  1995, AJ, 110, 463

\bibitem[\protect\citeauthoryear{{R Development Core Team}}{{R Development Core
  Team}}{2008}]{R}
{R Development Core Team} 2008, R: A Language and Environment for Statistical
  Computing.
R Foundation for Statistical Computing, Vienna, Austria

\bibitem[\protect\citeauthoryear{{Reisenegger}, {Quintana}, {Carrasco} \&
  {Maze}}{{Reisenegger} et~al.}{2000}]{Reisenegger00}
{Reisenegger} A.,  {Quintana} H.,  {Carrasco} E.~R.,    {Maze} J.,  2000, AJ,
  120, 523

\bibitem[\protect\citeauthoryear{{Reisenegger}, {Quintana}, {Proust} \&
  {Slezak}}{{Reisenegger} et~al.}{2002}]{Reisenegger02}
{Reisenegger} A.,  {Quintana} H.,  {Proust} D.,    {Slezak} E.,  2002, The
  Messenger, 107, 18

\bibitem[\protect\citeauthoryear{{Rood}}{{Rood}}{1976}]{Rood76}
{Rood} H.~J.,  1976, ApJ, 207, 16

\bibitem[\protect\citeauthoryear{{Saglia}, {Bender} \& {Dressler}}{{Saglia}
  et~al.}{1993}]{Saglia93}
{Saglia} R.~P.,  {Bender} R.,    {Dressler} A.,  1993, A\&A, 279, 75

\bibitem[\protect\citeauthoryear{Scholz}{Scholz}{2012}]{adk}
Scholz F.,  2012, adk: Anderson-Darling K-Sample Test and Combinations of Such
  Tests

\bibitem[\protect\citeauthoryear{{Shane} \& {Wirtanen}}{{Shane} \&
  {Wirtanen}}{1954}]{Shane54}
{Shane} C.~D.,  {Wirtanen} C.~A.,  1954, AJ, 59, 285

\bibitem[\protect\citeauthoryear{{Shapley}}{{Shapley}}{1930}]{Shapley30}
{Shapley} H.,  1930, Harvard College Observatory Bulletin, 874, 9

\bibitem[\protect\citeauthoryear{{Small}, {Ma}, {Sargent} \&
  {Hamilton}}{{Small} et~al.}{1998}]{small3}
{Small} T.~A.,  {Ma} C.-P.,  {Sargent} W.~L.~W.,    {Hamilton} D.,  1998, ApJ,
  492, 45

\bibitem[\protect\citeauthoryear{{Small}, {Sargent} \& {Hamilton}}{{Small}
  et~al.}{1997a}]{small1}
{Small} T.~A.,  {Sargent} W.~L.~W.,    {Hamilton} D.,  1997a, ApJS, 111, 1

\bibitem[\protect\citeauthoryear{{Small}, {Sargent} \& {Hamilton}}{{Small}
  et~al.}{1997b}]{small2}
{Small} T.~A.,  {Sargent} W.~L.~W.,    {Hamilton} D.,  1997b, ApJ, 487, 512

\end{thebibliography}
\end{document}